\documentstyle[12pt,epsfig]{article}
\def\be{\begin{equation}}
\def\bea{\begin{eqnarray}}
\def\ee{\end{equation}}
\def\eea{\end{eqnarray}}
                              \def\barr{\begin{array}}
                              \def\earr{\end{array}}
\def\dis{\displaystyle}

\def\etc{{\em etc.}}
\def\etal{{\em et al.}}

\def\lsim{\:\raisebox{-0.5ex}{$\stackrel{\textstyle<}{\sim}$}\:}
\def\gsim{\:\raisebox{-0.5ex}{$\stackrel{\textstyle>}{\sim}$}\:}
                               
                              \def\gev{\: \rm GeV} 
                              \def\tev{\: \rm TeV} 
                              
                              \def\fb {\: \rm fb}
\def\gappeq{\mathrel{\rlap {\raise.5ex\hbox{$>$}}
            {\lower.5ex\hbox{$\sim$}}}}
\def\lappeq{\mathrel{\rlap{\raise.5ex\hbox{$<$}}
            {\lower.5ex\hbox{$\sim$}}}}
\def\ra{\rightarrow}

\def\squ{\tilde q}
\def\slep{\tilde l}
\def\slashiii#1{\setbox0=\hbox{$#1$}#1\hskip-\wd0\hbox to\wd0{\hss\sl/\/\hss}}

\newcommand{\rpv}{$\slashiii{R}$}
\begin{document}                                                              
\begin{flushright}
MRI-PHY/P20000428 \\
{\large \tt hep-ph/0005082}
\end{flushright}

\vspace*{2ex}

\begin{center}
{\Large\bf Prospects of Squark \& Slepton searches at a Gamma
 Gamma Collider}
\vskip 15pt

{\sf Debajyoti Choudhury$^{1}$ and Anindya Datta${^2}$}

{\footnotesize Mehta Research Institute, 
Chhatnag Road, Jhusi, Allahabad 211 019, India. \\
E-mail: $^1$debchou@.mri.ernet.in, $^2$anindya@mri.ernet.in}\\

\vskip 15pt
{\Large\bf Abstract}
\end{center}

We examine the prospects of detecting sfermions at a gamma gamma
collider.  Once produced, a slepton  can decay into a pair of 
quarks (jets) through $R$-parity violating interactions. 
Similarly, a squark may decay into a lepton-quark pair. 
Analyzing the corresponding Standard Model backgrounds, namely 
4-jet and dilepton plus dijet final states respectively, we show 
that the sfermion can be detected almost right upto the kinematic 
limit and its mass determined to a fair degree of accuracy. Similar 
statements also hold for nonsupersymmetric leptoquarks and diquarks.
\vskip 20pt

PACS numbers: 12.60.Jv, 14.80.Ly, 13.88.+e

\vskip 20pt
\hspace*{0.65cm}

\def\baselinestretch{1.0}


\section{Introduction}
	\label{sec:introd}
The Standard Model (SM) of electroweak interactions, while eminently
successful in describing most available data, is rightly not regarded 
as the final theory. Apart from aesthetic drawbacks such as the 
lack of explanations for either the fermion masses or the relative strength 
of the gauge couplings, it also suffers from a few technical problems. To
overcome these lacunae, many authors have, over the years, proposed 
models going beyond the SM. Two of the most attractive classes of 
such models are those incorporating grand unification~\cite{GUT}
and/or supersymmetry~\cite{SUSY}. Both, as well as other models, 
predict additional particle states. In this article we shall concentrate 
on the scalar sector of such theories. 

Within the SM, baryon ($B$) and lepton ($L$) number conservation come 
about due to
accidental symmetries. In other words, such conservation is not guaranteed 
by any theoretical reasons, 
but are rather the consequences of the choice of the particle 
content\footnote{Indeed, nonperturbative effects within the SM itself 
	do break $B + L$ symmetry.}.
In extensions of the SM, such an accidental occurrence is obviously not 
guaranteed. For example, even in the simplest grand unified theories (GUTs), 
both the gauge and the scalar sector interactions
violate each of $B$ and $L$. The 
corresponding particles, namely the diquarks~\cite{diquarks} and 
leptoquarks~\cite{leptoquarks} have been studied in the literature to a 
considerable extent. Simultaneous breaking of both $B$ and $L$ symmetry
can be disastrous though as that would lead to rapid proton decay. Within
GUTs, gauge boson-mediated proton decay is naturally suppressed on account
of their large mass; in the scalar sector, the particle content 
can be chosen such that there is no diquark-leptoquark mixing, at least 
as far as the light sector is concerned.

In the case of the Minimal Supersymmetric Standard Model (MSSM), 
on the other hand, we do not have the option of demanding the 
`offending'  fields (the supersymmetric partners of the SM fermions)
to be superheavy. Ruling out the undesirable terms necessitates the 
introduction of a discrete 
symmetry, $R \equiv (-1)^{3 (B - L) + 2 S}$ (with $S$ denoting
the spin of the field)~\cite{fayet}.
Apart from ruling out both $B$ and $L$ violating terms in the superpotential,
this symmetry has the additional consequence of rendering the lightest 
supersymmetric partner absolutely stable. However, such a symmetry is 
{\em ad hoc}. Hence, it is of interest 
to consider possible violations of this symmetry especially since it 
has rather important experimental consequences, not the least of which 
concerns the detection of the supersymmetric partners.

As we shall see later, there are certain similarities between \rpv\ 
interactions on the one hand and leptoquarks and/or diquarks on the other. 
An example of this is afforded by the explanations~\cite{HERA_expl} 
of the anomalous large-$Q^2$ data reported by the {\sc Hera} 
collaborations. However, the \rpv-MSSM, being a richer (low-energy) theory, 
offers a larger set of possibilities, both in the context of the 
{\sc Hera} events as well as other anomalies like the ones observed 
at {\sc Karmen}\cite{karmen} or at {\sc Kamiokande}\cite{kamiokande}.
Hence, in our discussions, we shall concentrate primarily on the 
supersymmetric case, and point out, as special cases, the corresponding 
results for theories with leptoquarks or diquarks.

It is only natural that such particles would be looked for in existing 
colliders. Indeed, the best lower bounds on the masses of such particles 
have been obtained from analyses Tevatron data. Pair production 
of leptoquarks (or equivalently squarks decaying through an $L$ violating 
interaction) leads to a final state comprising a dilepton pair alongwith 
jets~\cite{Tev_dilep}. More interestingly, in the supersymmetric case,
gluino production cross-section is larger and, in addition,
can lead to like-sign dileptons, thereby making the signal stand out even 
more~\cite{Tev_likesign}. However, all such analyses, of necessity, make 
certain assumptions that are not necessarily true. Similar is the case 
of searches at {\sc Hera} which depend on the size of the $L$-violating 
coupling~\cite{H1_LQ}. To minimize the dependence on such assumptions, it 
is thus necessary to consider further experiments.

In this article we investigate the production of such scalars 
(supersymmetric or otherwise) at a photon collider. The 
lepton (or baryon) number violating decays may result in 
a significant excess in dijet plus dilepton or 4-jet final states. 
An analysis of such data would allow us not only to detect such particles 
but also to measure their masses and, to an extent, the branching 
fractions.
The plan of the rest of the article is as follows. In the next section we
discuss briefly the photon photon collider and the production
cross-section of these scalar fermions at a such a collider. In
section~\ref{sec:sfermion_decays}, 
we focus on the \rpv\  violating couplings and the decays of
the squarks and the sleptons. Section~\ref{sec:signal_bkgd}
will be devoted to comparing
the signals and the possible SM backgrounds. We will explore the
discovery/ exclusion limits on the SUSY parameter space using our
signal and background analysis in section~\ref{sec:bound_on_susy_param}. 
Finally, we summarise in section~\ref{sec:concl}.

\section{Pair-production of Scalars at a Photon Collider} 
	\label{sec:pair_prod}
To the leading order in perturbation theory, the cross-section 
for charged particle pair-production at a gamma gamma collider is
completely model independent
This is quite unlike the case for the $e^+ e^-$ machine where the 
coupling of the scalar to the $Z$ as well as its Yukawa couplings 
have a bearing on the answer. Thus,  model dependence appears
only in the decay channels, and hence is easier to analyse.
Let us consider, first, the case of monochromatic photon beams. If the 
center-of-mass energy be $\sqrt{s}$,  and the 
product of the photon helicities 
(circular polarization)\footnote{We do not consider here the possibility 
		of linear polarization. The most general expression can be 
		found in Ref.~\protect\cite{CCGM}.}
be $P_{\gamma \gamma}$, the pair-production cross-section for a scalar 
of charge $Q$ and mass $m$ is given by
\be 
\begin{array}{rcl}
\dis
\frac{ {\rm d} \sigma}{ {\rm d} \Omega} 
	& =  & \dis
          \frac{Q^4 N_c \alpha^2}{ s} \; \beta \;
           \Bigg[ \left(1 + P_{\gamma \gamma} \right)
		  \: \Big\{ \frac{1}{2} 
			   - \frac{ \beta^2 \sin^2 \,\theta}
				  {1 - \beta^2 \cos^2 \;\theta}
		     \Big\}
		+ \frac{ \beta^4 \sin^4 \,\theta}
				  {(1 - \beta^2 \cos^2 \;\theta)^2}
 		\Bigg]  \ .
\end{array}
	\label{cs:fixed_ener}
\ee
In eq.(\ref{cs:fixed_ener}), 
$\beta \equiv (1 - 4 m^2 / \hat s)$ is the velocity of 
the scalars in the center of mass frame
and $\theta$ the corresponding scattering angle. The colour factor 
$N_c = 1(3)$ for sleptons (squarks). The
unpolarised cross-section can easily be obtained from the above
expression by setting $P_{\gamma \gamma} =$ 0.  One can check
easily that,  for small scalar masses, the dominant mode is the 
$P_{\gamma \gamma} = -1$ mode. On the other hand, the
$P_{\gamma \gamma} = 1$ mode dominates for scalar masses 
near the kinematic limit~\cite{CCGM}.

In reality though, high energy monochromatic photon beams are 
extremely unlikely. In fact, the only known way to obtain 
very high energy photon beams is to induce laser back-scattering 
off an energetic $e^\pm$ beam~\cite{telnov}. The reflected 
photon beam carries off only a fraction ($y$) of the $e^\pm$ 
energy with 
\be
\barr{rcl}
y_{\rm max} & = & \dis \frac{z}{1 + z} 
	\\[2ex]
z & \equiv & \dis 
	\frac{4 E_e E_L}{m_e^2} 
                 \cos^2 \frac{\theta_{e L}}{2}
\earr
\ee
where $E_{e (L)}$ are the energies of the incident $e^\pm$ beam 
and the laser respectively and $\theta_{e L}$ is the incidence angle.
One can, in principle, increase the photon energy by increasing the 
energy of the laser beam. However, increasing $E_L$
also enhances the probability of electron positron
pair creation through laser and scattered-photon interactions, and 
consequently results in beam degradation. 
An optimal choice of $z$ taking care of this is $z = 2(1 + \sqrt{2})$,
and this is the value that we adopt in our analysis.

The cross-sections for a realistic photon-photon collider 
can then be obtained by convoluting the fixed-energy 
cross-sections of eq.(\ref{cs:fixed_ener}) with the appropriate 
photon spectrum. For circularly polarized lasers scattering off 
polarised electron beams, the number-density $n(y)$ and average 
helicity $\xi(y)$ for the scattered photons are given by~\cite{telnov}
\be
\barr{rcl}
\dis \frac{dn}{dy} &=&  \dis 
	\frac{2 \pi \alpha^2}{m_{e}^2 z \sigma_C} C(y) 
   \\[2ex]
\xi (y) &=& \dis 
	\frac{1}{C(y)} \bigg[ P_e \bigg\{ \frac{y}{1-y} + y(2 r -1)^2
\bigg \} - P_l (2 r -1) \bigg( 1 - y + \frac{1}{1-y} \bigg) \bigg] 
    \\[2ex]
C(y) &\equiv& \dis 
	\frac{y}{1-y} + (1 -y) - 4r(1-r) - 2P_e P_l rz (2r -1)(2 -y)
\earr
\ee
where $r \equiv y / z / (1 - y)$ and $\sigma_C$ is the total 
Compton cross-section. In the study of polarized beams, one fact 
needs to be borne in mind. While full (100\%) polarisation is possible
for a laser, it is unlikely to be realized for electrons. In the rest of
this article we shall consider the electron polarization, wherever 
applicable, to be 90\%.

\begin{figure}[htb]
\centerline{
\epsfxsize=7.6cm\epsfysize=7.0cm
                     \epsfbox{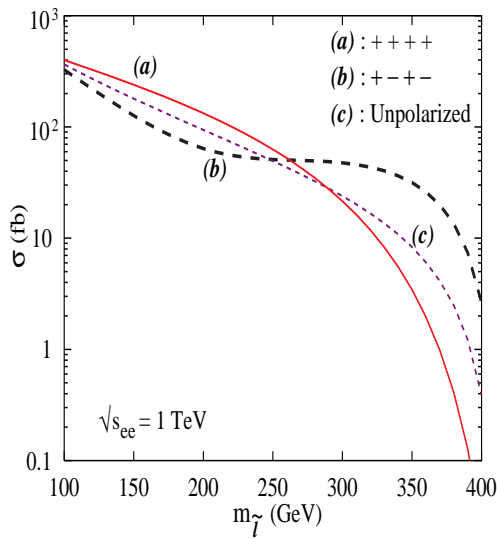}
}
\caption{\em Cross-section for slepton production at a $\gamma \gamma$
collider }

\label{fig:prod}
\end{figure}

In Fig.\ref{fig:prod},
we present the total cross-section for slepton pair
production at a photon collider wherein the parent $e^+ e^-$ 
(or $e^- e^-$) collider operates at a center of mass
energy of $1 \tev$. We present the results for three 
combinations of incident laser and electron
polarizations ($L_1 e_1 L_2 e_2$). 
For the entire mass range, at least one of the
two polarised ($++++$ and $+-+-$) cross-sections wins
over the unpolarised case. 
When scalar masses are less than $230 \gev$, cross-section for all three
cases are comparable with 
$\sigma_{++++} > \sigma_{\rm unp} > \sigma_{+-+-}$.
But for scalar masses above $230 \gev$ this hierarchy is just the opposite.
In this region, $\sigma_{+-+-}$ falls off more slowly than the other two.
Depending on $m_{\tilde f}$, $\sigma_{+-+-}$ can be 5--8 times larger than 
$\sigma_{\rm unp}$ in this mass range.     

\section{R-parity Violating Decays of the Sfermions.}
	\label{sec:sfermion_decays}
In this section we present a very brief overview of the phenomenology 
of $R$-parity violation within the MSSM with particular emphasis on 
the additional decay channels available to the sfermions. 
As has been noted in the literature, unless a discrete 
symmetry
is introduced explicitly, the superpotential,  in addition to the normal 
Yukawa terms, can also contain the following terms:
\be
{\cal W}_{R\!\!\!/} = \mu_i L_i H_2  
                    + \lambda_{ijk} L^i_L L^j_L \bar E^k_R
	            + \lambda'_{ijk} L^i_L Q^j_L \bar{D}^k_R.
	            + \lambda''_{ijk} \bar{U}^i_R  \bar{D}^j_R \bar{D}^k_R.
	\label{superpot}
\ee
In the above, while $L^i_L$ and $Q^i_L$ denote the left-handed
lepton and quark doublet superfields
respectively, $E^i_R, U^i_R$ and $D^i_R$ denote
the corresponding right-handed superfields. The couplings $\lambda_{ijk}$ 
are antisymmetric under the exchange of the first two indices, while 
$\lambda''_{ijk}$ are antisymmetric under the exchange of the 
last two. Since the 36 couplings $\lambda_{ijk}$ and 
$\lambda'_{ijk}$ violate lepton-number while the other 9 ($\lambda''_{ijk}$)
violate baryon number, simultaneous existence of both the sets of operators
would lead to a rapid proton decay and is hence strongly disfavoured. 
We will thus consider either $L$ violating or $B$ violating couplings.
Furthermore, even within one such subset, non-zero values of more than 
one coupling could lead to large flavour-changing neutral current 
amplitudes~\cite{fcnc}. We will thus restrict ourselves to cases where 
only one such coupling is 
dominant\footnote{This assumption, of course, prevents us from considering 
	certain spectacular collider signatures.}.

Examining the individual couplings {\em vis a vis} direct \rpv\   
decays of sfermions, we easily notice:
\begin{itemize}
	\item the couplings $\lambda_{ijk}$ connect either a charged slepton 
	      to a lepton-neutrino pair or a sneutrino to two charged leptons.
	      While sneutrino pair production is irrelevant in the context 
	      of photon colliders, charged sleptons will lead to a final 
	      state with a lepton pair and missing energy-momentum, a state 
	      with a large SM background emanating from $W$-pair production;
	\item the couplings $\lambda'_{ijk}$ lead to both squark-quark-lepton
	      and slepton-quark-quark vertices. While the first resembles 
	      a leptoquark vertex, the second (apart from colour factors) 
	      mimics a diquark vertex (although there is no violation 
	      of baryon number);
	\item the couplings $\lambda''_{ijk}$ lead to squark-quark-quark 
	      vertices, and again mimic diquarks as far as direct \rpv\   
	      decays are concerned.
\end{itemize}
Thus direct decays through $\lambda_{ijk}$ are of no concern to us. 
Since we shall not address the question of cascading decays in this 
article, we do not consider such couplings any further. Similarly, as far 
as direct decays are concerned, the phenomenology of $\lambda''_{ijk}$
is very similar to that of slepton pair production and subsequent decay
through some $\lambda'_{ijk}$. Hence it suffices to consider the 
case of a single non-zero $\lambda'_{ijk}$. Analogous results for 
$\lambda''_{ijk}$ can easily be deduced from those that we present. 

Expressed in terms of the component fields, the 
relevant part of the  Lagrangian reads 
\be
\barr{rcl}
{\cal L}_{\lambda'} & = & \dis 
  \lambda'_{ijk} \bigg[ \,
   {\tilde \nu}^i_L {\bar d}^k_R d^j_L
 + {\tilde d}^j_L {\bar d}^k_R \nu^i_L
 + ({\tilde d}^k_R)^\ast ({\bar \nu}^i_L)^c d^j_L   
         \\ 
&& \hspace*{2em} \dis 
	- {\tilde e}^i_L {\bar d}^k_R u^j_L
   - {\tilde u}^j_L {\bar d}^k_R e^i_L
   - ({\tilde d}^k_R)^\ast ({\bar e}^i_L)^c u^j_L \,\bigg]
 + h.c
\earr
\ee
Bounds on these couplings can be obtained from various low-energy
observables~\cite{bounds}. These include, for example, meson decay 
widths~\cite{meson_decay}, neutrino masses~\cite{neutrino_mass}, 
rates for neutrinoless double beta decay~\cite{bb0nu} {\em etc}. 
The bounds generally scale with the sfermion mass, and for 
$m_{\tilde f} = 100 \gev$ range from $\sim 0.02$ to $0.8$. 

%
\begin{figure}[ht]
\centerline{
\epsfxsize=6.5cm\epsfysize=6.0cm
                     \epsfbox{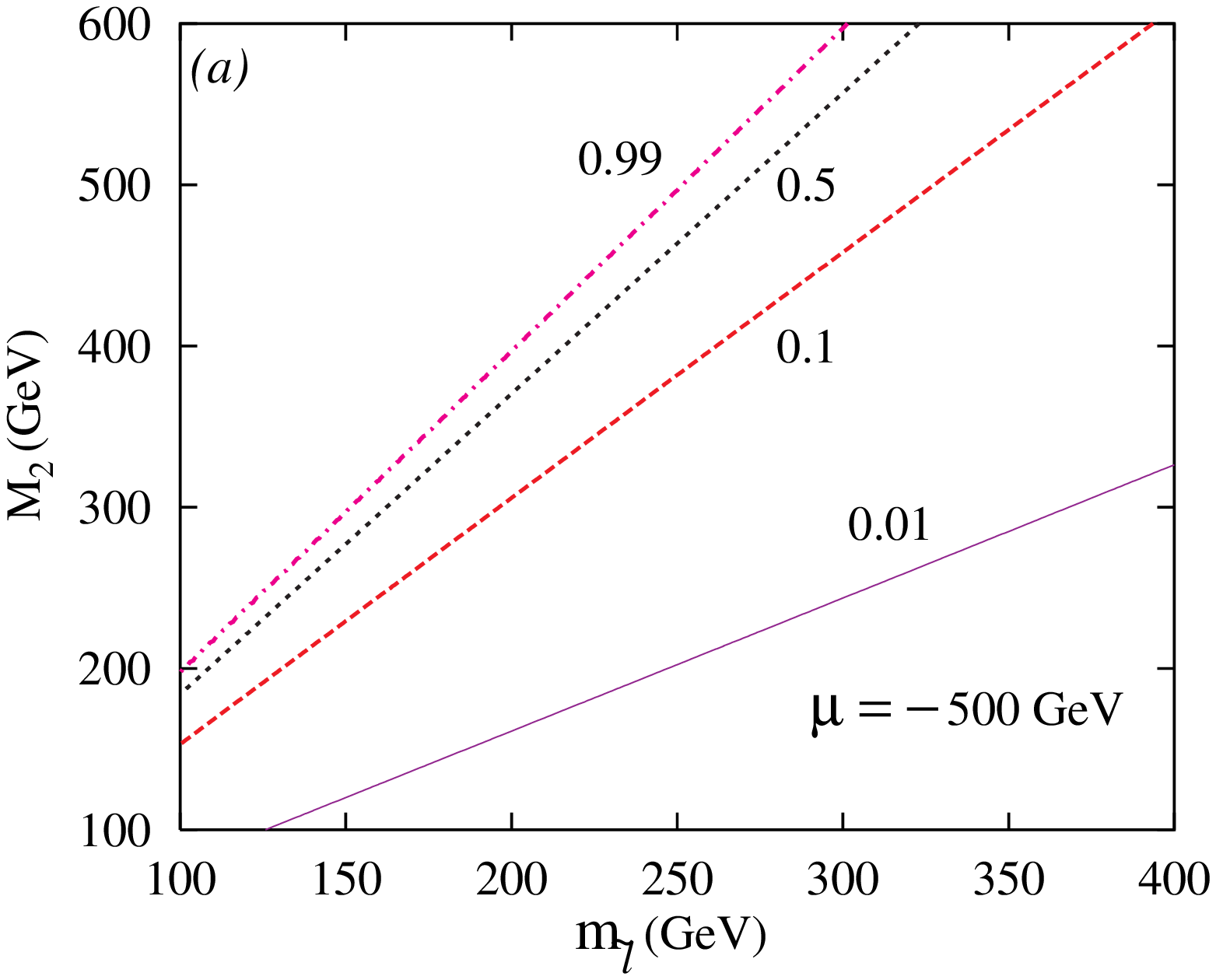}
	\hspace*{-5ex}
\epsfxsize=6.5cm\epsfysize=6.0cm
                     \epsfbox{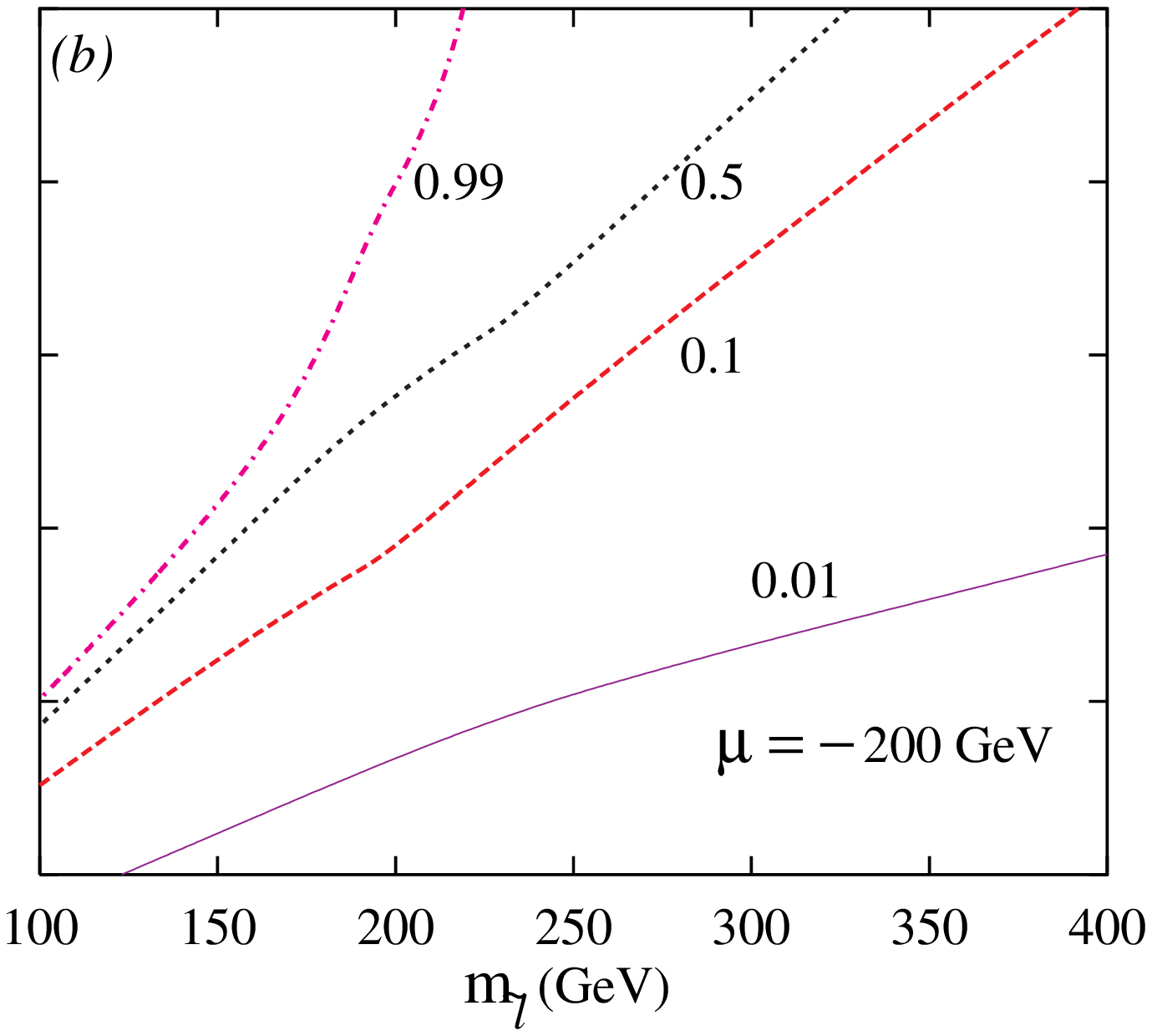}
}

\vspace*{4ex}
\centerline{
\epsfxsize=6.5cm\epsfysize=6.0cm
                     \epsfbox{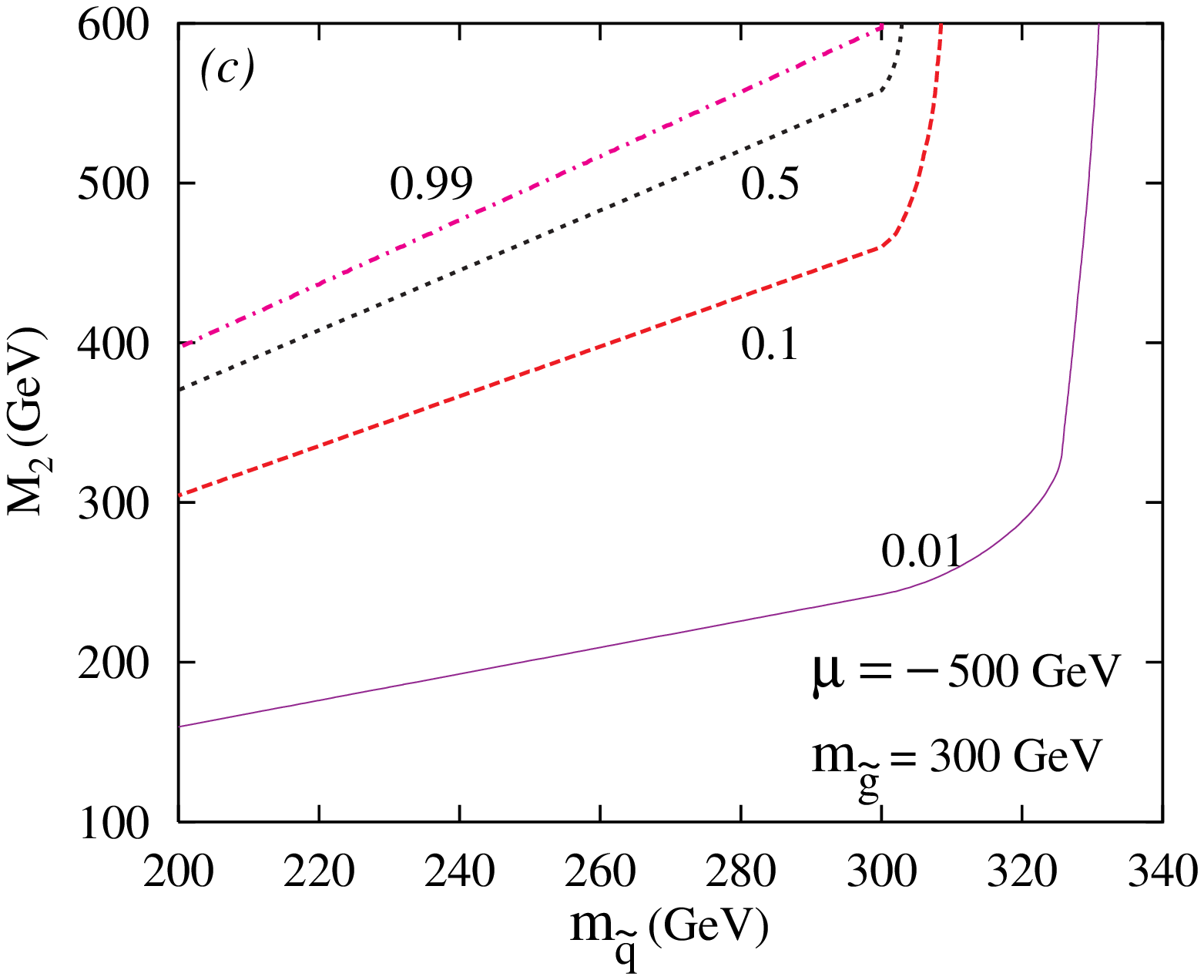}
	\hspace*{-5ex}
\epsfxsize=6.5cm\epsfysize=6.0cm
                     \epsfbox{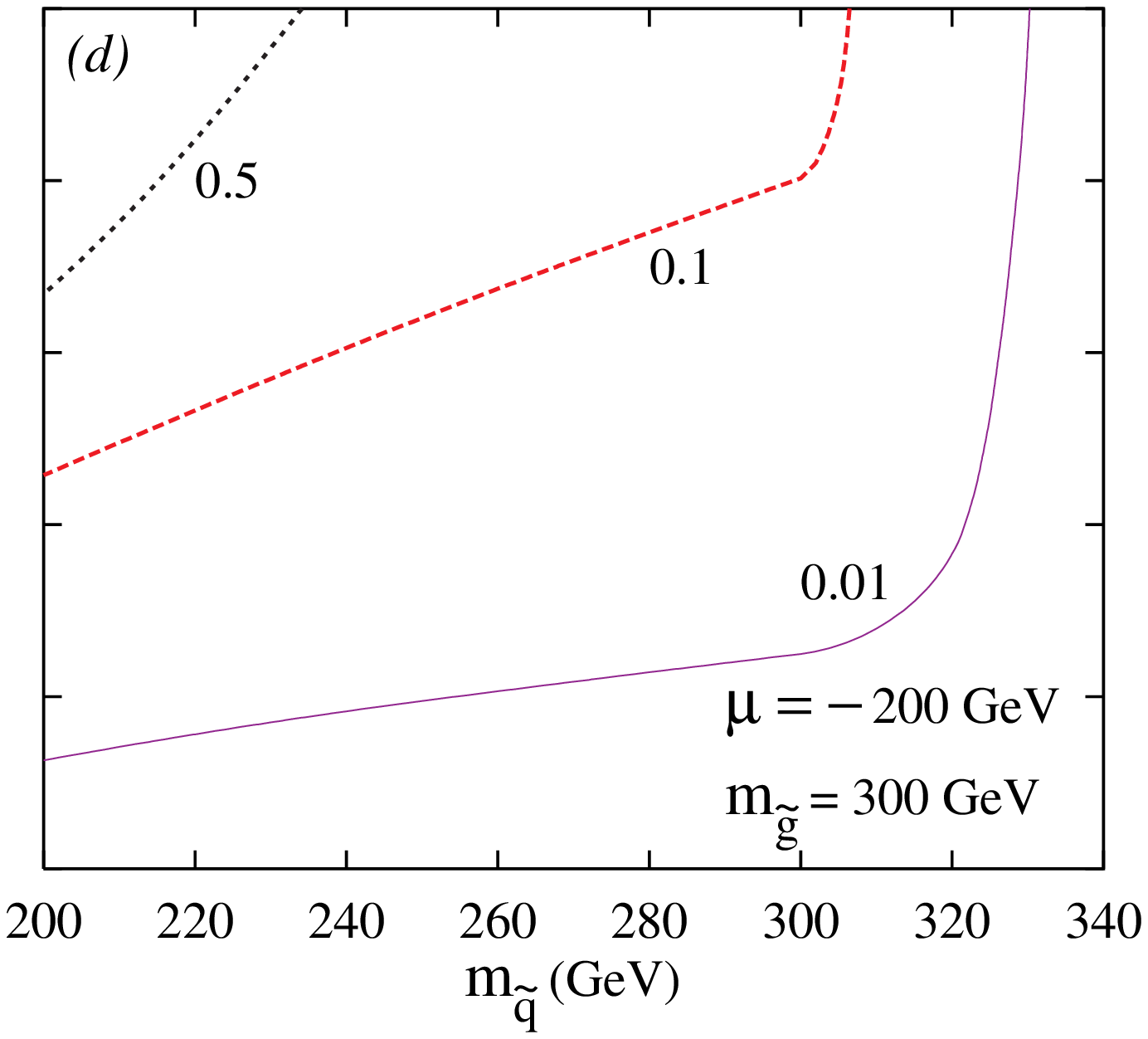}
}
\caption{\em Contours for constant \rpv- decay branching ratios. 
    {\em (a)} slepton with $\mu = -500 \gev$, 
    {\em (b)} slepton with $\mu = -200 \gev$, 
    {\em (c)} squark with $\mu = -500 \gev$ and
    {\em (d)} squark with $\mu = -200 \gev$. 
In each case, $\lambda' = 0.02$, $\tan\beta = 5$. For squark decays, we have 
assumed the gluino mass to be $300 \gev$. 
}
\label{fig:br}
\end{figure}
In presence of such a \rpv\ term, a sfermion can decay into two 
SM fermions. This mode then competes with the $R$-conserving ones 
namely $\tilde f \ra f + \chi_i^0, \: f' + \chi_i^\pm$. The partial 
decay widths for the latter are of course determined by the masses 
of the neutralinos (charginos) and their couplings to the sfermion.
These quantities, in turn, are determined by the gaugino mass 
parameters $M_i$, the higgsino mass $\mu$ and the ratio of the 
two higgs vacuum expectation values $\tan \beta$. However, if 
decays into the top quark are not allowed, the dependence on 
the last two parameters is negligible. Furthermore, if we assume 
gaugino mass unification, as happens in GUTs or supergravity 
inspired scenarios, only one of the parameters $M_i$ is independent. 
In Fig.\ref{fig:br}, we plot the
contours of constant \rpv-branching ratios for 
sleptons and up-type squarks in $M_2$-$m_{\tilde f}$ 
plane. The particular value of $\lambda'_{ijk}$ chosen here 
satisfies the strongest (barring the case of $\lambda'_{111}$) 
of the low energy constraints for $m_{\tilde f} \gsim 100 \gev$.
To reduce the number of parameters, 
we assume the GUT relation between the $U(1)$ and $SU(2)$ gaugino
masses but retain the gluino mass as a free parameter. The lower 
limit on the squark mass has been chosen so as to be 
consistent\footnote{In actuality, their bound is somewhat larger 
	than $200 \gev$, but has been obtained for a special case.}
with the bounds quoted by the CDF collaboration~\cite{Tev_dilep}.

Let us concentrate first on Fig.\ref{fig:br}$a$, where we consider the 
case of a slepton and $\mu = -500 \gev$. The higgsino mass being 
large, two of the neutralinos (as well as one chargino state) 
are irrelevant over the entire range of 
parameter space displayed. The slepton may only decay into the two lighter 
neutralinos (mainly gauginos) and the lighter chargino.
The straight lines thus reflect contours of 
equal kinematic suppression in the decay widths. The curves change 
somewhat for a smaller value of $\mu$ (see Fig.\ref{fig:br}$b$) since now 
the neutralino (chargino) sector mixings ensure that more decay 
channels are available to the slepton (especially if it is on the heavier 
side). 

For smaller masses, the contours for the squarks
(Figs.\ref{fig:br}$c,d$) are qualitatively similar. The main
difference arises on account of the gluino, which we assume, for the
purpose of the graph, to be significantly lighter than that stipulated
by gaugino mass unification.  Since the decay into gluino proceeds
through strong interactions, it dominates almost immediately on 
crossing the kinematic threshold. It must be pointed out though that,
for the stop, this decay (or, for that matter, the
decays into neutralinos) almost never reaches the kinematic threshold
in the parameter range of interest.

\section{Signals and Backgrounds}
	\label{sec:signal_bkgd}
We are now in a position to discuss the signals that we are 
interested in and the backgrounds thereof. As
already discussed, we focus on the direct \rpv--decays of the 
sfermions. Thus, the sleptons decay into two quarks each, while 
for the squarks we concentrate on $l (e, \mu) +
jet$. Thus, the pair produced charged scalars will give
rise to $l^+ l^-$ plus 2 jet or 4 jet in the final state.

The obvious background consists of the SM process 
$\gamma \gamma \ra 4 \ {\rm fermions}$. This has contributions from both 
`resonant' (such as $W^+ W^-$ production) and nonresonant diagrams. 
Another potential source is 
heavy quark ($b,c$) production followed by decays of these quarks. 
However, in general such decay products are soft and such backgrounds 
can be eliminated by imposing simple kinematical cuts. 
Also to be considered are the contributions from both
single-- and double--resolved processes. 
These turn out to be negligible though.
The large number of diagrams contributing to the background are
calculated using the helicity amplitude package {\sc
madgraph}~\cite{MadGraph}.  To estimate the number of events and their
distribution(s), we use a parton-level Monte-Carlo event
generator. 

\subsection{Squark Production}
	\label{sec:squark}
We begin with the squarks as the analysis is simpler. To be specific, 
we consider only the up-type ($Q = 2/3$) squarks as the production 
cross-section is 16 times than that for the down-type squarks. The only 
direct \rpv\   decay channel is therefore into a charged lepton and a 
down-type quark with the resultant final state consisting of
two hard jets alongwith two hard leptons. 

For given quark and lepton flavours, there are {\em forty} SM diagrams 
contributing to the the process $\gamma \gamma \ra \ell^+ \ell^- q \bar q$. 
These can be divided into three topological classes:
\begin{itemize}
   \item 12 of the type $\gamma \gamma \ra \ell^+ \ell^- \gamma^* (Z^*)$
	 with the off-shell boson emanating from any of the three lepton 
	 lines and subsequently going into the quark pair;
   \item 12 of the type $\gamma \gamma \ra q \bar q \gamma^* (Z^*)$;
   \item 16 diagrams with a `$t$-channel' $\gamma (Z)$ exchange (nonresonant
	 topology).
\end{itemize}
Clearly, for small angle scatterings, each of these diagrams can lead 
to a large contribution. To eliminate these, we require that both jets 
and leptons should be relatively central:
\be
|\eta_{j,\ell}| < 2.5 \ .
	\label{cuts:eta}
\ee
This also ensures that these would be well within the detector. Clearly,
the loss in signal would be marginal as the final state there arises 
from decay of two scalar particles, with even the production cross-section 
not being strongly peaked. 
We also must ensure that the jets and the leptons are separated well
enough to be detectable as individual entities. To this end, we 
adapt the well-known cone algorithm for jet separation to a
parton-level analysis. Defining 
$\Delta R_{ab} \equiv \sqrt{ (\Delta \eta_{ab})^2 + (\Delta \phi_{ab})^2 }$
where $\Delta \eta$ is the difference of the rapidities of the two 
particles and $\Delta \phi$ is their azimuthal separation, we demand that
\be
\Delta R_{jj} \geq 0.7 \ , \qquad 
\Delta R_{jl} > 0.5 \ , \qquad 
\Delta R_{ll} > 0.2 \ .
	\label{cuts:R}
\ee 
Detectability also requires that these particles (jets) must have a 
minimum momentum. Over and above this, it should be noted that the signal
events would typically be characterized by all the four particles having 
relatively large transverse momenta ($p_T$). 
On the other hand, the SM background
has a large component wherein at least one fermion pair has relatively 
small $p_T$. A cut on the particle momenta is thus called for. We find 
that rather than imposing the same requirement on all the particles, it 
is better to order them (leptons and jets individually)
according to their $p_T$. However, in doing this, one must 
take into account the detector resolution effects. We incorporate
this into our analysis by means of a rather conservative 
smearing of 
energies\footnote{The expected angular resolutions are too fine 
		  to be of any concern to us.}~\cite{Mur_Pes}:
\be
\barr{rclcl}
\dis \frac{\delta E_j}{E_j}  
       & = & \dis \frac{0.4}{\sqrt{E_j / 1 \gev} } + 0.02 
       & \qquad & {\rm for \ jets} \ ,
                 \\[2.5ex]
\dis \frac{\delta E_\ell}{E_\ell}  
       & = & \dis \frac{0.15}{\sqrt{E_\ell / 1 \gev} } + 0.01
       & \qquad & {\rm for \ leptons} \ .
\earr
	\label{resolution}
\ee
We then demand that 
\be
p_T^{j1} , p_T^{\ell 1}  > 25 \gev, \quad 
p_T^{j2} , p_T^{\ell 2} > 20 \gev, 
	\label{cuts:pT:squark}
\ee
where $j1$ denotes the jet with larger transverse momentum \etc.
Alongwith the separability requirement (eq.\ref{cuts:R}), this 
also serves to eliminate the bulk of contributions from the 
$\gamma \gamma \ra \ell^+ \ell^- \gamma^*$ and 
$\gamma \gamma \ra q \bar q \gamma^*$ diagrams.

As we have discussed before, the background also contains diagrams 
of the form $\gamma \gamma \ra f \bar f Z^*$ with the $Z$ going into
the other pair of fermions. Eliminating these necessitates that we 
discard events where either the lepton-lepton or the jet-jet 
invariant mass is close to $m_Z$:
\be
	m_{\ell \ell}\ , m_{j j} \not \in [80 \gev, 100 \gev] \ .
	\label{cuts:Zpole:squark}
\ee
Imposing the above set of cuts, we can then calculate both the signal 
and background. For the latter, we need to sum over all possible 
light quarks\footnote{We do not consider here the special case that the 
	squark decays into a lepton and $b$-quark thus making it possible for 
	us to tag the corresponding jet.}. 
Doing this, we obtain, for unpolarized beams and $\sqrt{s_{ee}} = 1 \tev$, 
an integrated cross-section of $\sigma_{\rm Bkgd} = 2.32 \fb$. 
To further improve the signal-to-noise ratio, we can use a 
particular feature of the signal event topology. 
In the limit of infinite momentum resolution,
we can, for each of these events, 
find one particular lepton--jet pairing such that the  
corresponding invariant masses are exactly the same. Provided such 
a pairing is unambiguous, this invariant mass is then the mass of the 
squark. Clearly, the background events would not show the same 
characteristics, and this can be used to our advantage. 
We then retain only those events for which 
\be
	\left| M^{(1)}_{\ell j}  - M^{(2)}_{\ell j} \right| 
	  \leq 10 \gev
   \label{cuts:massrecon:squark}
\ee
for at least one such pairing. The corresponding {\em average} mass 
is then treated as our determination of $m_{\tilde q}$. 
With this additional restriction, the background cross-section drops to 
$\sigma_{\rm Bkgd} = 0.14 \fb$. 
It might be argued, and rightly too, 
that a $m_{\tilde q}$ independent criterion as in 
eq.(\ref{cuts:massrecon:squark}) is not the most efficient one. 
Indeed, with the 
fractional energy resolution growing with the jet/lepton energies, we would 
do better by optimizing this cut for each squark value that we might be 
interested in. However, we omit to do so in order to keep the analysis 
a simple one. 

\begin{figure}[htb]
\centerline{
\epsfxsize=6.5cm\epsfysize=6.0cm
                     \epsfbox{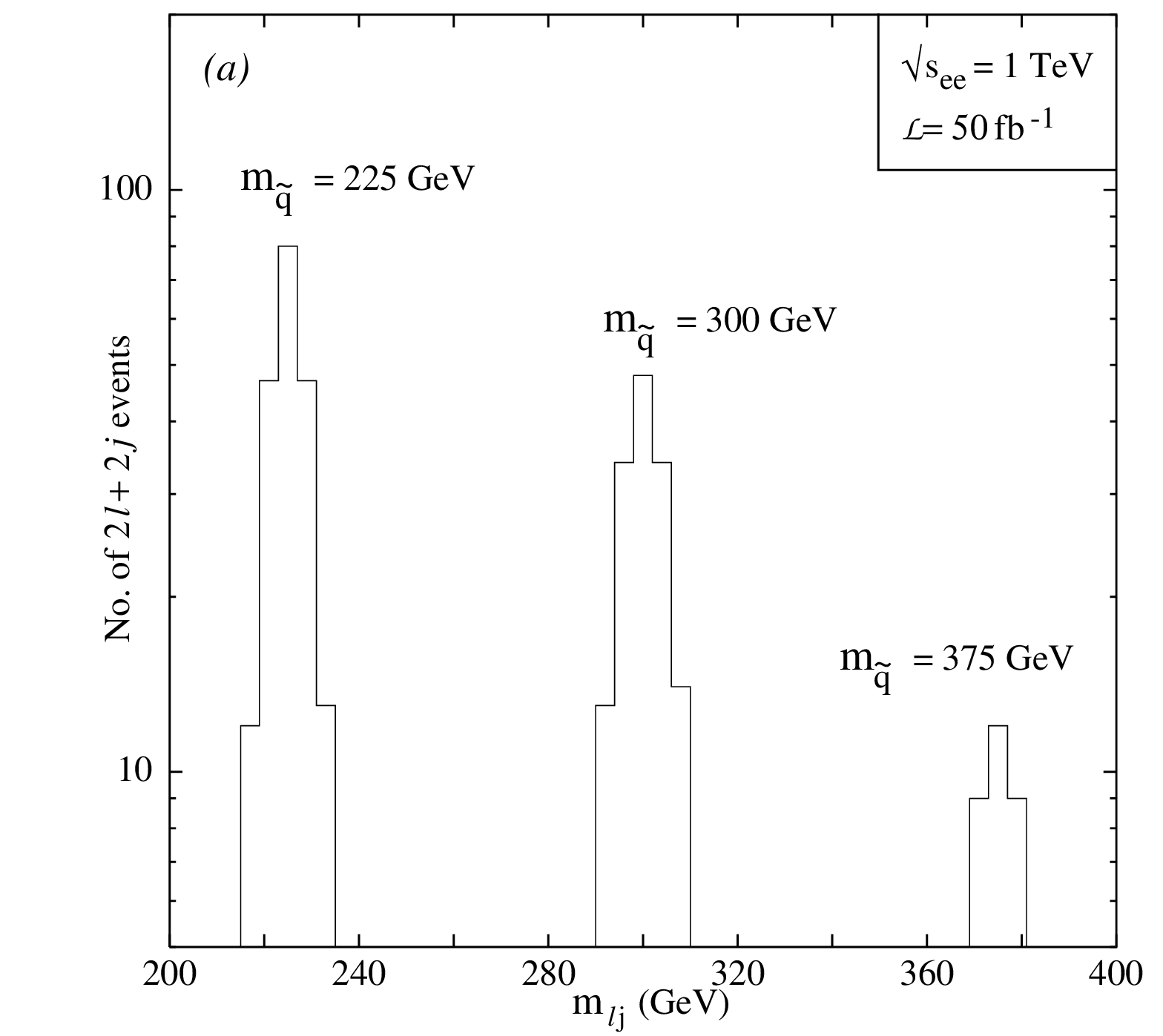}
\epsfxsize=6.5cm\epsfysize=6.0cm
                     \epsfbox{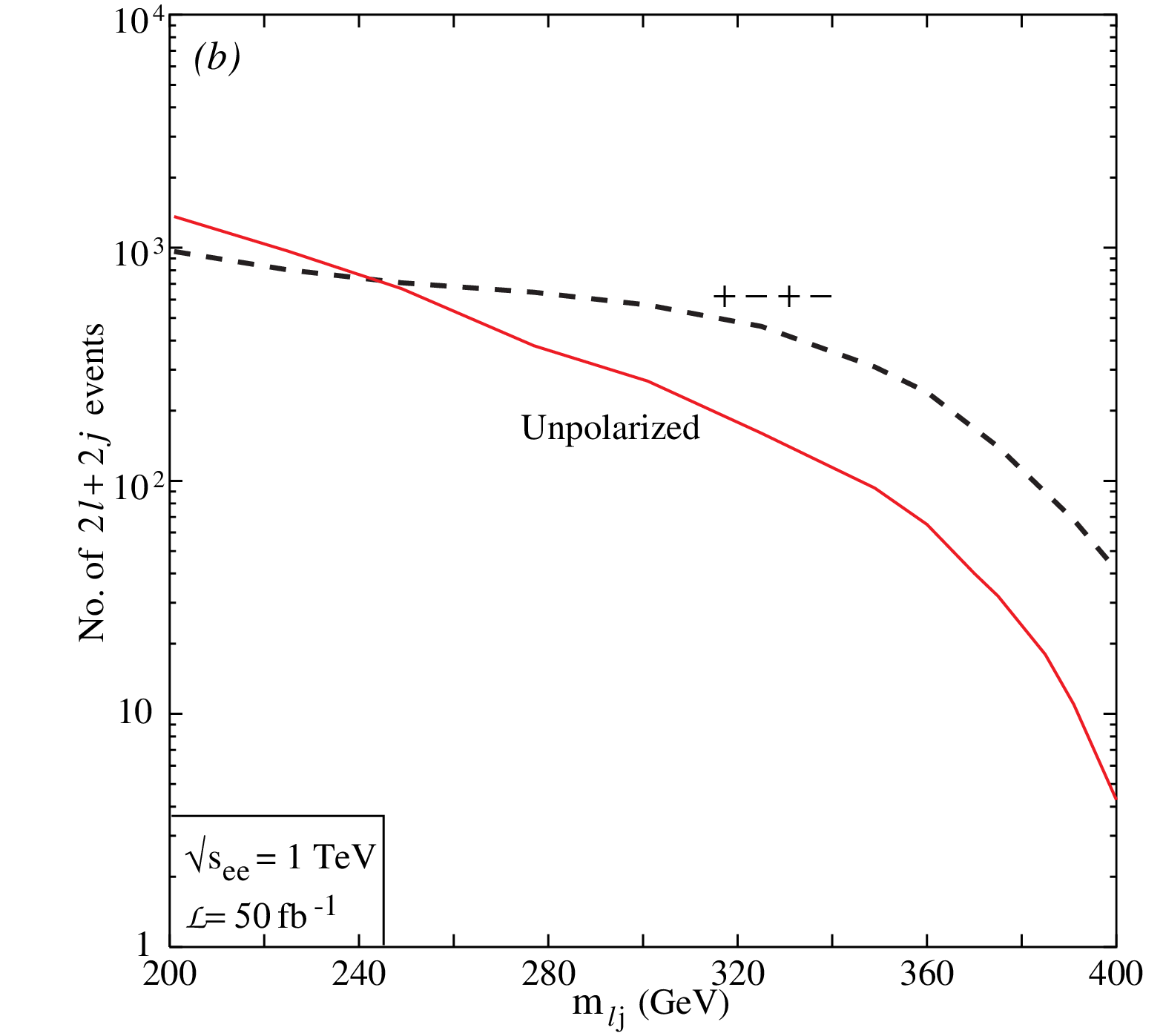}
}
\caption{\em {\em (a)} Mass reconstruction from the dilepton + dijet 
final mode at a photon-photon collider with the parent machine 
operating at center of mass energy of $1 \tev$. 
The cuts of eqs.(\ref{cuts:eta}--\ref{cuts:massrecon:squark})
have been used. The SM background is much smaller than the scale of the 
graph. 
{\em (b)} Number of  dilepton + dijet events due to squark 
production for both polarized and unpolarized initial beams.
}  \label{fig:sqrk_evts}
\end{figure}

With the above set of kinematical cuts, we have been able to reduce
the background to insignificant levels. It remains to be seen how much
of the signal is retained and whether mass reconstruction is
possible. We tackle the questions in the reverse order. The
impediments to mass reconstruction come from two sources. The first is
of course the effect of the resolution smearing. A second source of
ambiguity is the possibility that both set of reconstructions could
satisfy eq.(\ref{cuts:massrecon:squark}).  In such a case, we retain the
pairing with the {\em larger} average ($M_{\rm av}$) of the
reconstructed masses.  As is easily ascertained from 
Fig.\ref{fig:sqrk_evts}$a$, the mass
reconstruction works quite well for relatively smaller values of
$m_{\tilde q}$. For heavier squarks though,
the peaks are not as sharp. This feature is easy to understand. As
$m_{\tilde q}$ increases, the squarks are produced with smaller and
smaller average momentum.  Close to the kinematic threshold, the mass
difference (eq.\ref{cuts:massrecon:squark}), in the limit of infinite
resolution, would vanish identically for {\em both} the pairings. 
Of course, if the resolution was really infinite, our algorithm---namely 
the larger of the two reconstructed masses---would still make the 
correct identification. However, because of the resolution smearing, 
migration of events do occur. Thus, sensitivity could be 
increased were we to compare,
bin by bin, the data and the SM expectations, thereby using a
statistical discriminator (such as a $\chi^2$ test). In this article,
though, we will not attempt such an analysis. Rather, we effect the
somewhat cruder strategy of integrating over five contiguous 
bins\footnote{As the lepton and quark momenta
are smeared using a Gaussian distribution, the signal invariant mass
distribution has not a sharp peak in the relevant mass bin.  Rather it
also shows a Gaussian structure. To take into account this 
smearing effect, when calculating the number of signal events for a
particular squark or slepton mass, we not only consider the number of
events in the relevant mass bin but also add to it the contribution
from the four adjacent bins.
}
centering around $M_{\rm av}$ and ascribing any excess therein to a
squark of  mass $M_{\rm av}$. The partial loss of information inherent in
such a procedure makes our results to be somewhat conservative. In
Fig.\ref{fig:sqrk_evts}$b$,  we present the number of events
expected (as a function of $m_{\tilde q}$) after the imposition of all
of the above cuts (we assume here 100\% branching into a lepton and 
a quark). Comparing it to Fig.\ref{fig:prod}, we see that the
loss in signal, while significant, is not debilitating.  On the other
hand, hardly any background events are expected within a given
bin (the surviving background cross-section of $0.14 \fb$ corresponds 
to just 7 events over the entire mass range for the assumed 
luminosity of $50 \fb^{-1}$). 
Consequently, observation of even an handful of such events, 
concentrated within a small mass range,  could be
construed as the evidence for squark production and subsequent decay
through an $R$-parity violating interaction, albeit with smaller
branching fractions. 

\begin{figure}[ht]
\centerline{
\epsfxsize=7.0cm\epsfysize=7.0cm
                     \epsfbox{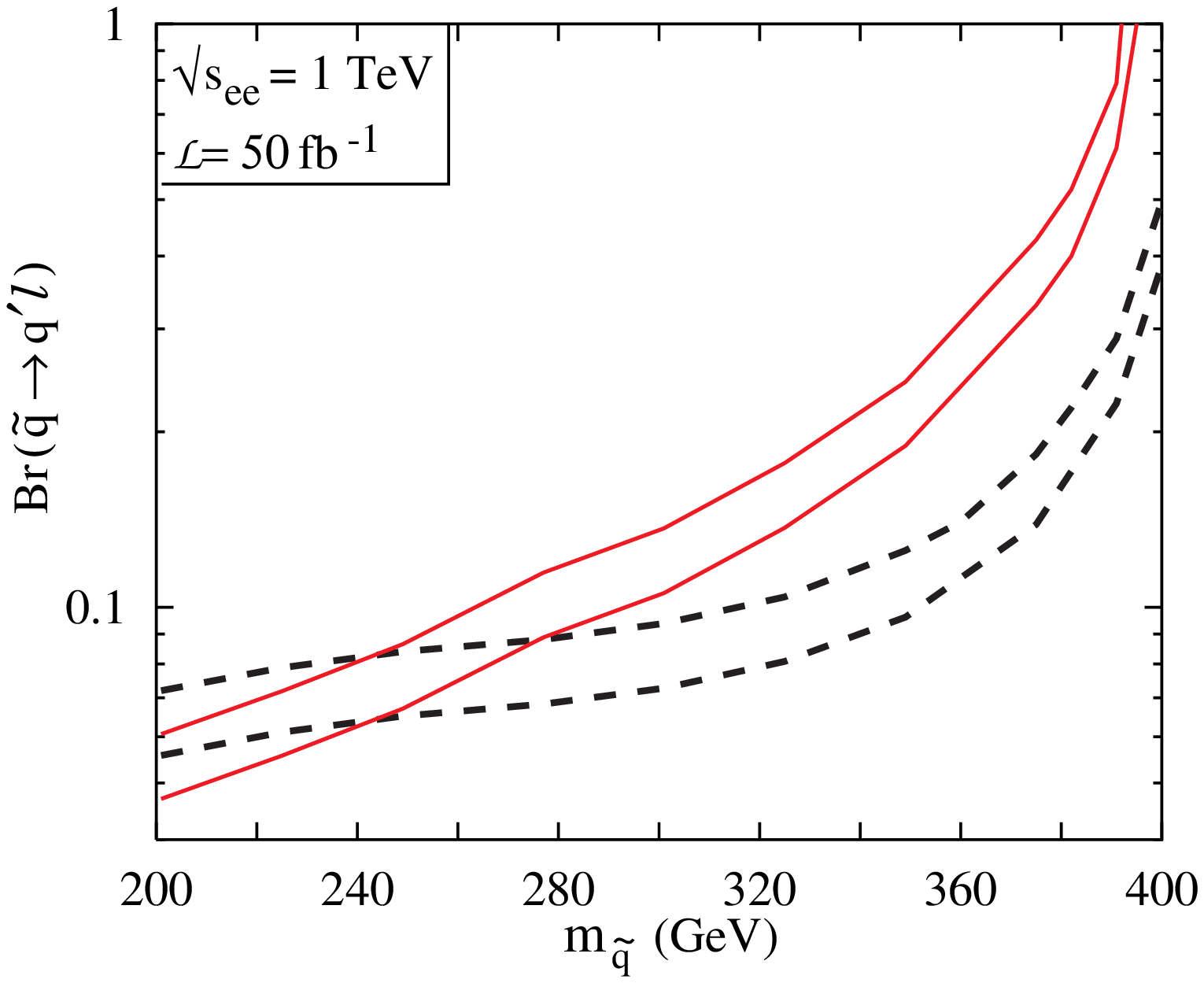}
}
\caption{\em Minimum branching ratio for the decay $\squ \ra l j$ 
	required for discovering (upper curve) or ruling out at 
	95\% C.L. (lower curve) an up-type squark.
	The solid and dashed lines correspond to unpolarized and
	$(+ - + -)$ polarized initial states respectively.  
	}
\label{fig:sqrk_excl}
\end{figure}

As we have commented in Section~\ref{sec:sfermion_decays}, 
in the supersymmetric case, the
squark may decay through $R$-conserving interactions into a quark 
and a chargino/neutralino. Although the latter will ultimately decay 
into SM particles through \rpv\ interactions, the event shape would be 
considerably different from that we have considered here. Confining 
ourselves solely to the analysis of 4-fermion final states, we can 
use the information of Fig.\ref{fig:sqrk_evts}$b$ to obtain 
exclusion/discovery limits in the $m_{\tilde q}$--branching fraction plane. 
Since the number of background events is almost zero, 
the required branching fraction $Br$ is approximately given by 
\[
{\rm Br} = \sqrt {\frac{n _{\rm req}}{N _s}}
\]
where $N_s$ is the number of signal events corresponding to 
${\rm Br} = 1$ and $n _{\rm req} = 5 \: (3)$ for discovery 
(95\% exclusion). In Fig. \ref{fig:sqrk_excl} we present the 
corresponding contours 
for two different choices of the initial state polarisation. 
As the polarised $(+-+-$) cross-section dominates over the unpolarised 
for large $m _{\squ}$, the required branching ratios are consequently smaller.
For example, close to the kinematic limit (say $m_{\squ} \approx 400$), 
we would make a discovery even with a branching fraction 50\% 
provided we work with the correct beam polarisation. 
On the other hand, with an unpolarised initial state,
and $m _{\squ} \gsim 390 \gev$, 
signal events number less than five even for a 100\% branching.

\subsection{Slepton Production}
	\label{sec:slepton}
The production cross-section for charged sleptons is higher 
(by a factor of $27/16$) than that for $up$-type squarks. 
However, since they decay into two quarks each, the resultant 
final state of 4 jets is more complicated than that in the previous 
subsection. In Table.~\ref{table:4j}, we list the various subprocesses
contributing to this background.
\begin{table}[htb]
\begin{center}
\begin{tabular}{|c|l|c|c|c|}
\hline
\hline
   No &	\multicolumn{1}{|c|}{Subprocess type}
                   & \multicolumn{2}{|c|}{Number of diagrams}
                   & \multicolumn{1}{|p{0.8in}|}{Number of subprocesses}
       \\[-0.7ex]
     \cline{3-4}
    &&&&   \\[-1ex]
    &   & ${\cal O}(\alpha \alpha_s)$  & ${\cal O}(\alpha^2)$ & 
    \\[0.5ex]
\hline     
1. &
  $\gamma \gamma  \ra u_i \bar u_i g g $ 
            & 30 &  0  &  2 \\[0.6ex]
2. &
  $\gamma \gamma  \ra d_i \bar d_i g g $ 
            & 30 &  0  &  3 \\[0.6ex]
\hline
3. &
  $\gamma \gamma  \ra u_i \bar u_i u_i \bar u_i $ 
            & 40 &  80  &  2 \\[0.6ex]
4. &
  $\gamma \gamma  \ra d_i \bar d_i d_i \bar d_i $ 
            & 40 &  80  &  3 \\[0.6ex]
\hline
5. &
  $\gamma \gamma  \ra u \bar u c \bar c $ 
            & 20 &  40  &  1 \\[0.6ex]
6. &
  $\gamma \gamma  \ra d_i \bar d_i d_j \bar d_j $ ($i \neq j$)
            & 20 &  40  &  3 \\[0.6ex]
\hline
 &&&& \\[-1.5ex]
7. &
  $\gamma \gamma  \ra u_i \bar u_i d_i \bar d_i $ 
            & 20 &  71  &  2 \\[0.6ex]
\hline
 &&&& \\[-1.5ex]
8. &
  $\gamma \gamma  \ra u_i \bar u_i  d_j \bar d_j $ ($i \neq j$)
            & 20 &  40  &  4 \\[0.6ex]
\hline
 &&&& \\[-1.5ex]
9. &
  $\gamma \gamma  \ra u \bar c d_i \bar d_j $ ($i \neq j$)
            & 0 &  31  &  3 \\[0.6ex]
\hline
\hline
\end{tabular}
	\end{center}  
\caption{\em Number of Feynman diagrams for various leading-order subprocesses
	     contributing to $\gamma \gamma \ra 4 \: {\rm jets}$. $i, j$ 
	     are the family indices.  }
	\label{table:4j}
\end{table}


Despite the profusion of diagrams and the attendant complications, 
certain kinematical cuts suggest themselves. Drawing from the experience 
of the previous subsection, we demand that
\be
|\eta_{j}| < 2.5  \ ,
	\label{cuts:eta:slepton}
\ee
and
\be
\Delta R_{jj} \geq 0.7 \ .
	\label{cuts:R:slepton}
\ee
An important feature is that the background receives 
${\cal O}(\alpha \alpha_s)$ contributions and that many more 
subprocesses contribute to it as compared to that in the previous 
subsection. Consequently, the background is much larger and we 
need to impose somewhat stricter $p_T$ requirements. Once again ordering 
the jets by their transverse momentum, we demand that
\be
p_T^{j1} , p_T^{j2}  > 40 \gev, \qquad
 p_T^{j3} , p_T^{j4} > 15 \gev \ . 
	\label{cuts:pT:slepton}
\ee
This helps us to eliminate the bulk of the ${\cal O}(\alpha \alpha_s)$
contributions. The ${\cal O}(\alpha^2)$ contributions, on the other 
hand, are dominated by the resonant contributions. In the present 
case, these are of two types : 
  ($i$) $\gamma \gamma \ra f \bar f Z^*$ as before, and
  ($ii$) $\gamma \gamma \ra W^* W^*$. 
To eliminate both sets, we demand that 
\be
	m_{i k} \not \in [75 \gev, 95 \gev]
	\label{cuts:poles:slepton}
\ee
for any of the six pairings. Comparing it to the analogous 
cut of the last subsection, it might seem that a somewhat harder cut 
is called for. However, such a course would  entail the loss 
of a significant fraction of the signal as well. Unless we design 
$m_{\tilde \ell}$-specific cuts, eq.(\ref{cuts:poles:slepton}) 
was found to be an optimal choice. With these set of selection 
criteria, the processes of Table.~\ref{table:4j} lead to a total 
of $\sigma_{\rm Bkgd} \approx 170 \fb$ (for $\sqrt{s_{ee}} = 1 \tev$). 

\begin{figure}[hb]
\centerline{
\epsfxsize=7.0cm\epsfysize=6.5cm
                     \epsfbox{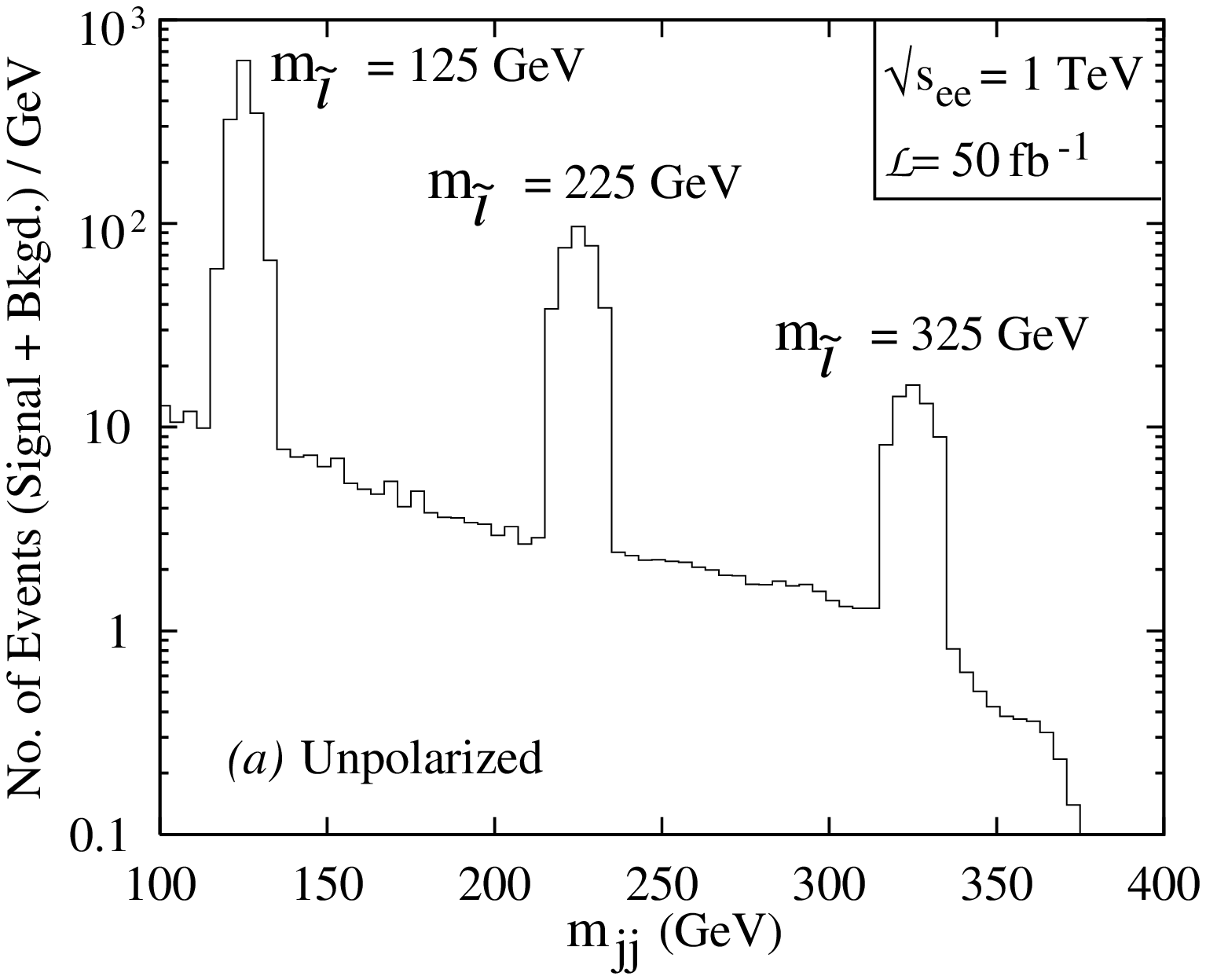}
	\hspace*{-7ex}
\epsfxsize=7.0cm\epsfysize=6.5cm
                     \epsfbox{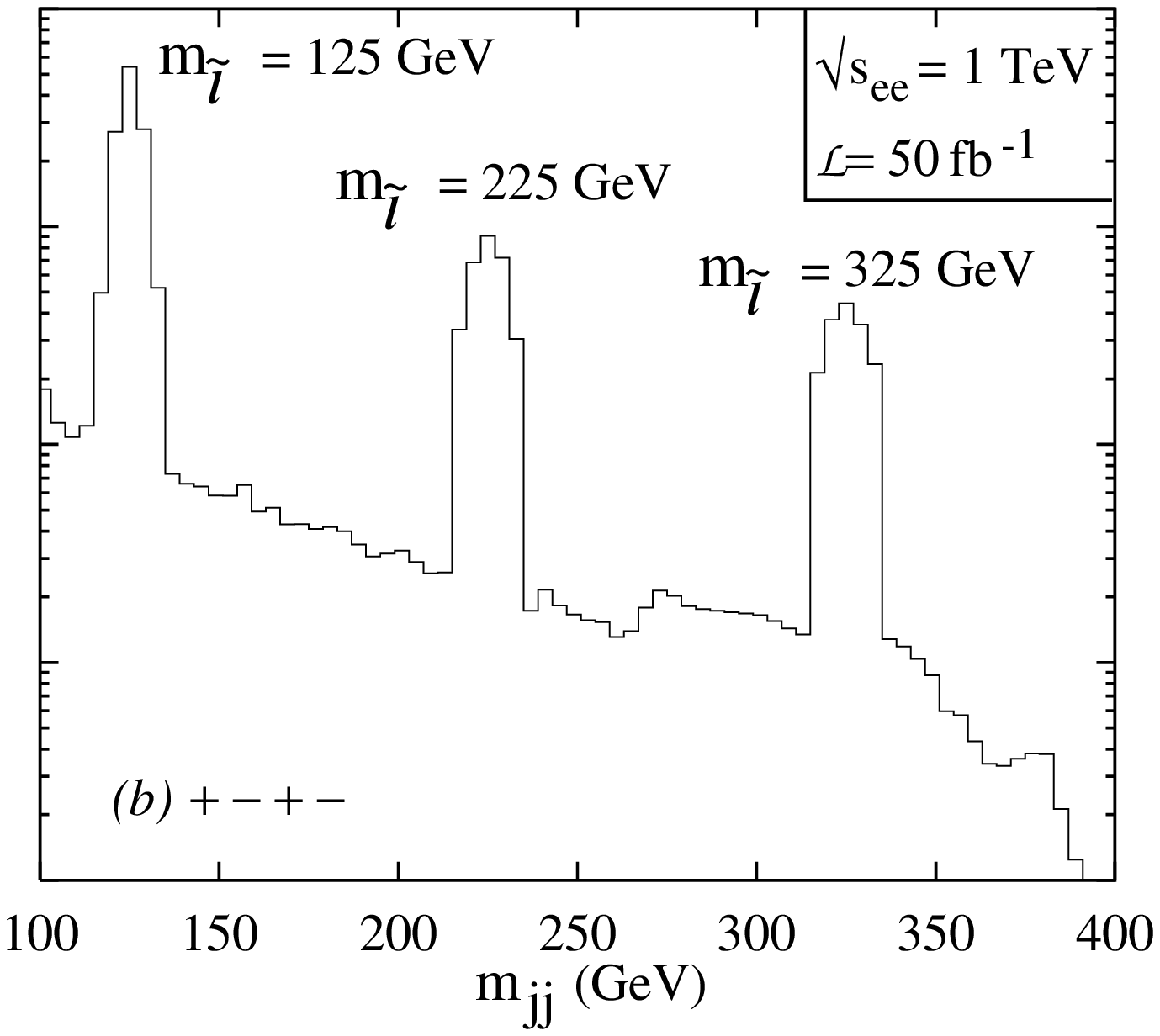}
}
\caption{\em Number of 4-jet events arising at a photon collider where 
the cuts of eqs.(\ref{cuts:eta:slepton}--\ref{cuts:massrecon:slepton}) 
have been imposed. The energy resolution is assumed to be the one given 
in eq.(\ref{resolution}). Thge continuum corresponds to the SM background 
and the peaks to sleptons decaying into two quarks each.
{\em (a)} is 
for unpolarised beams and {\em (b)} for polarised ($+-+-$) beams.}
\label{fig:slepton_evts}
\end{figure}
%
As in the case of the squark, we again take recourse to mass 
reconstruction, demanding that 
\be
	\left| M_{i j}  - M_{k l} \right| 
	  \leq 10 \gev
   \label{cuts:massrecon:slepton}
\ee
for at least one of the three pairings. Clearly, the combinatorial factor
for the SM background is higher in this case than that for squark decays. 
Consequently, the reduction ($\sigma_{\rm Bkgd} \ra\approx 20 \fb$) 
is not as pronounced. However, with most of this being concentrated at 
relatively small $M_{\rm av}$ (see Figs.\ref{fig:slepton_evts}a,b), 
where the signal cross-section is 
on the higher side too, even this  background is not a major source 
of concern. Also shown in Figs.\ref{fig:slepton_evts} are the signal 
profiles for particular values of $m_{\tilde l}$. The mass reconstruction 
is slightly worse here than that in the previous subsection. This 
is not unexpected as ($i$) jet energy resolution is significantly 
worse than lepton energy resolution; and ($ii$) here three different 
jet pairings are possible as compared to only two pairings in the other case.
Still, there is a significant excess of signal events over the background. 
Since the latter is rather insensitive to beam polarization, 
a careful choice of the same can be used to further enhance  this excess.
An example of this is provided by Fig.\ref{fig:slepton_evts}$b$.

\begin{figure}[ht]
\centerline{
\epsfxsize=7.6cm\epsfysize=7.0cm
                     \epsfbox{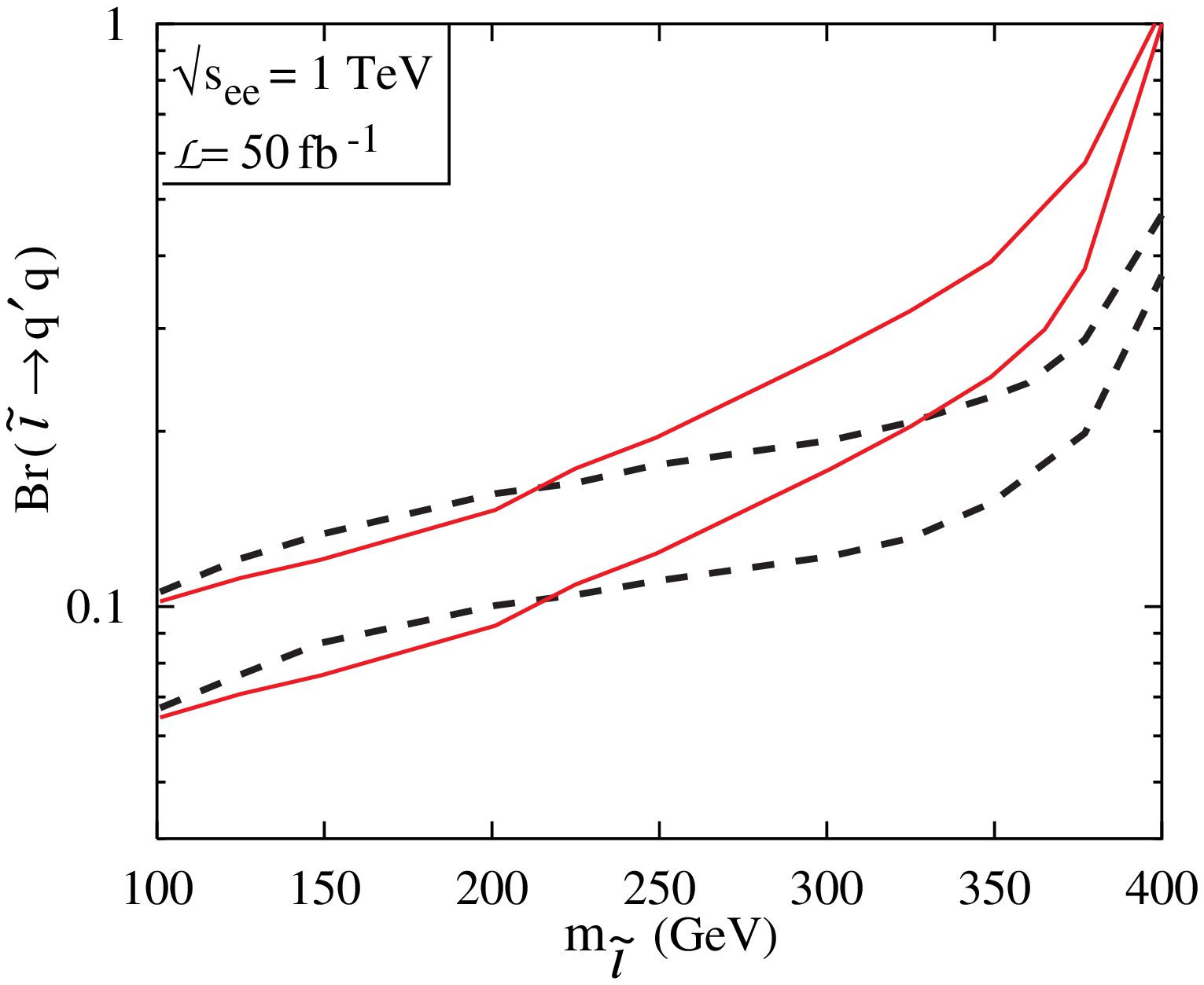}
}
\caption{\em Minimum branching ratio for the decay $\slep \ra q q'$ 
	required for discovering (upper curve) or ruling out at 
	95\% C.L. (lower curve) a slepton.
	The solid and dashed lines correspond to unpolarized and
	$(+ - + -)$ polarized initial states respectively.}
	\label{fig:slep_excl}
\end{figure}

To maximize sensitivity, we could either opt for mass-dependent kinematical 
cuts or compare events bin by bin and employ a statistical discriminator
such as a $\chi^2$ test. However, once again, we  adopt the simpler 
course of summing over five contiguous bins centered around the slepton 
mass of interest and compare it with the background. Since the latter 
is no longer vanishingly small, we cannot simplify our analysis as in 
the previous subsection. Instead, the required branching fraction is now 
given by
\[
Br. = \sqrt {\frac{n \sqrt{N _b}}{N _s}}
\]
where $N _b$ and $N _s$ are the  number of background and signal events 
(summed over the 5 bins). $n = 5 \: (2)$ for discovery (95 \% C.L. 
exclusion). Of course, this algorithm is valid strictly in the large
$N_b$ limit; for small $N_b$, we use the appropriate Poisson limit.
In Fig.\ref{fig:slep_excl}, we present the contours for two different 
polarization choices. For small sfermion masses, this channel 
is less sensitive compared to the one considered in the previous 
subsection, inspite of the larger production cross-section. This of course 
can be ascribed primarily to the much larger background and to a 
smaller extent to the combinatorial problem. For larger sfermion masses 
though, the background count is $\lsim {\cal O}(1)$ for both cases. 
Hence the exclusion curves are quite similar in this region. Of course, 
with a larger integrated luminosity, this argument would cease to hold
and the squark production channel
would outperform the present one even for $m_{\tilde f}$
close to the kinematic limit.

\subsubsection{Sleptons decaying into $b$ quarks}
	\label{sec:slepton_b}
The above results can be significantly improved if the slepton were to
decay into a $b$-quark and a light ($u$-type) quark. 
Since $b$-jets can be distinguished from those coming from light 
quarks or gluons, the number of processes contributing to 
the SM background is reduced considerably. Looking at Table~\ref{table:4j},
we see that, in the limit of ideal 
identification~\footnote{This approximation is quite valid given that
	misidentification probability  $\lsim 0.005$ 
	even in a hadronic environment~\protect\cite{Tev-2000}.},
only one subprocess each of types (2 \& 4) and two subprocesses each 
of types (6 \& 8) may contribute. Consequently, the background, with the 
an identical set of cuts, is now reduced to $\lsim 0.1 \fb$. 

\begin{figure}[htb]
\centerline{
\epsfxsize=6.5cm\epsfysize=6.0cm
                     \epsfbox{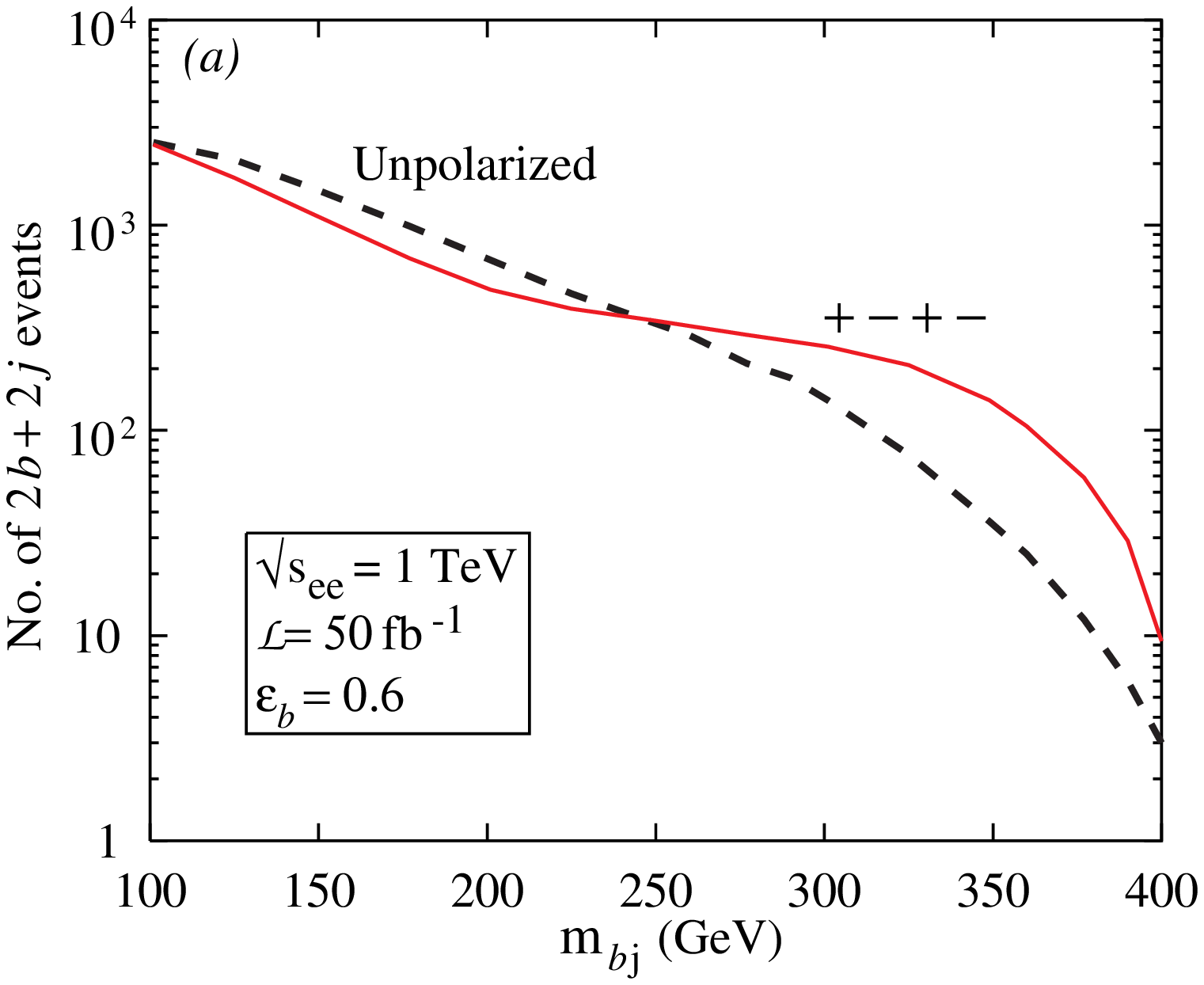}
	\hspace*{-2ex}
\epsfxsize=6.5cm\epsfysize=6.0cm
                     \epsfbox{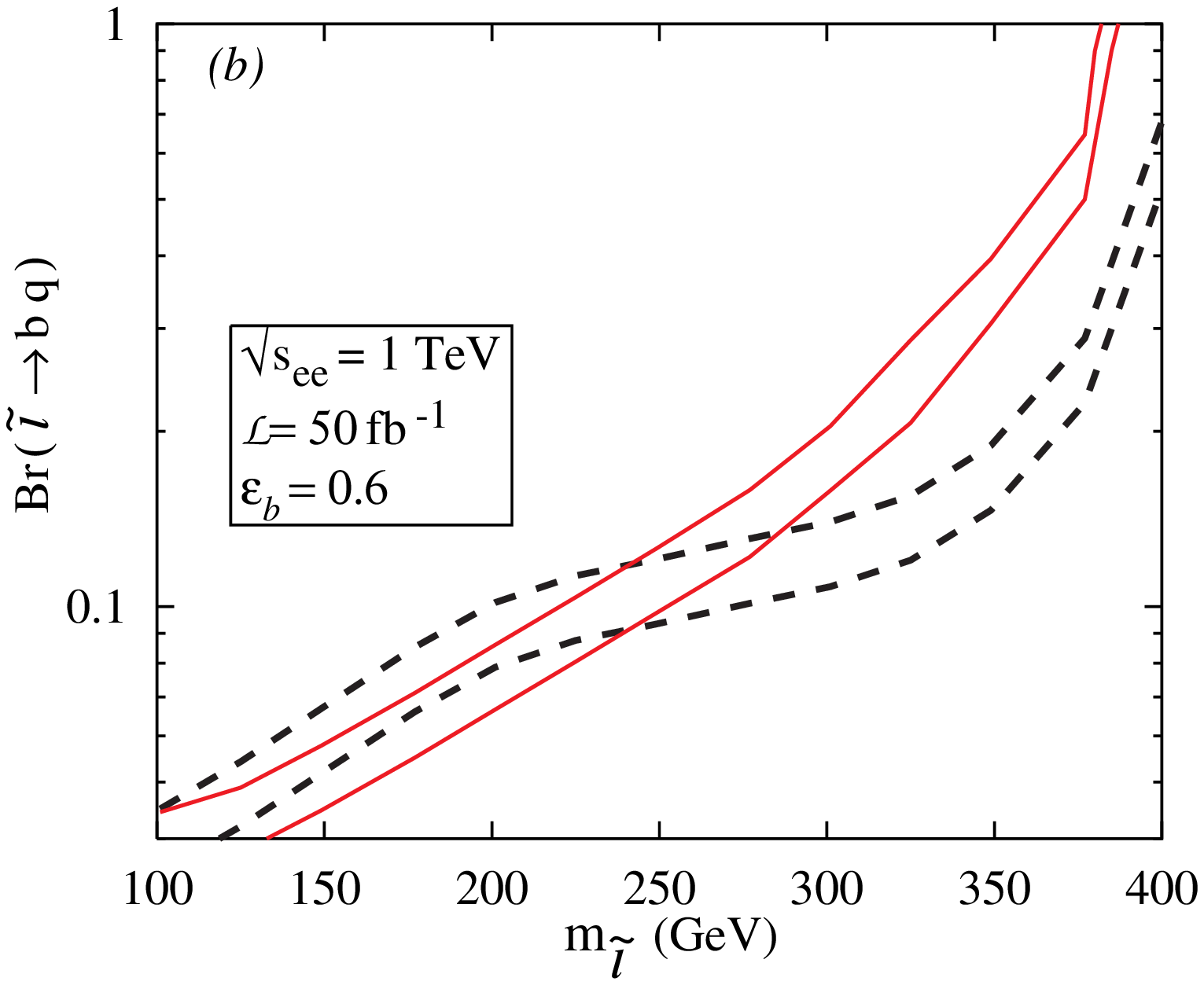}
}
\caption{\em {\em (a)} The number of $2b$ - $2j$ events coming from slepton 
production and decay; and 
	     {\em (b)} the minimum branching ratio for the decay 
	$\slep \ra b q$ 
	required for discovering (upper curve) or ruling out at 
	95\% C.L. (lower curve) a slepton.
	The solid and dashed lines correspond to unpolarized and
	$(+ - + -)$ polarized initial states respectively.
	}
\label{fig:slep_b}
\end{figure}

This enormous reduction in the background more than offsets the
reduction in the signal on account of the less-than-ideal $b$-tagging
efficiency. For the latter, we use the conservative value $\epsilon_b
= 0.6$ per $b$-jet~\cite{Tev-2000}. An additional improvement occurs
in the mass reconstruction on account of the smaller number of
combinatorial possibilities. The situation is thus quite analogous to
that of squark production, albeit with a smaller effective production
cross-section (the ratio being $\sim 27 \epsilon_b^2 / 16$).  That the
curves in Fig.\ref{fig:slep_b}$a$ are not exactly parallel to those of
Fig.\ref{fig:sqrk_evts}$b$ is due to the facts that the kinematic cuts
are not exactly the same and that the energy resolution is slightly
worse in the present case. The same also explains the shapes of the
exclusion/discovery curves in Fig.\ref{fig:slep_b}$b$. 

Interestingly, despite the smaller backgrounds, $b$-tagging does not 
seem to help much for large slepton masses (compare the curves of 
Fig.\ref{fig:slep_excl} and Fig.\ref{fig:slep_b}$b$). This is easily 
understood on noticing the fact that the SM background is essentially 
zero for such invariant masses. Consequently, the sensitivity of the 
channel is determined by the signal size alone. With a less-than-ideal 
efficiency, $b$-tagging obviously reduces the signal size without 
gaining in terms of the background. Foregoing tagging would improve 
the efficiency (for large $m_{\tilde l}$) to the levels of 
Fig.\ref{fig:slep_excl}, but at the cost of determining the nature 
of the coupling.

\section{Bounds on the SUSY Parameter Space}
	\label{sec:bound_on_susy_param}
In the previous sections we have presented the reach of a 
photon-photon collider in a {\em model independent} way. In other words, 
we have not made any assumptions about the decay modes 
allowed to these scalars. Once the rest of the spectrum in the theory
(and their couplings with the scalar in question) is known, the 
branching fractions to a pair of SM fermions can then be computed 
in terms of the Yukawa coupling (as, for example, was done 
in section~\ref{sec:sfermion_decays}). The branching ratios can 
then be combined with the exclusion/discovery plots
in a straightforward manner to yield the bounds on the relevant 
parameter space of the theory. For the sake of completeness, we 
present the outcome of one such exercise here. 

\begin{figure}[ht]
\centerline{
\epsfxsize=7.6cm\epsfysize=7.0cm
                     \epsfbox{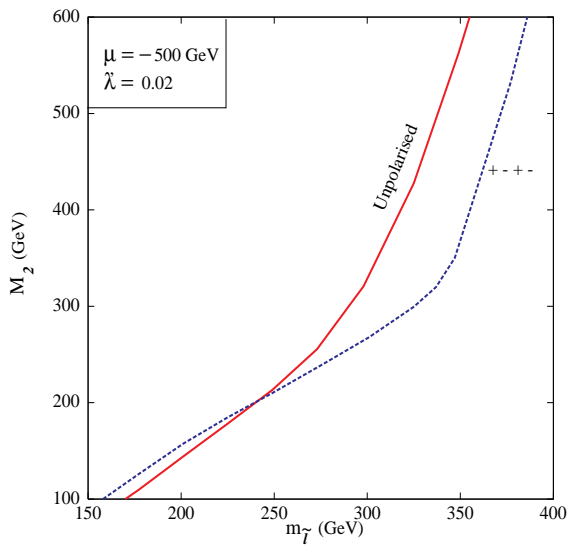}
}
\caption{\em Discovery contour in the $M _2$ - $m _{\tilde l}$ 
in the $2b$- $2j$ channel for a (unpolarised) and b(polarised) case.}
	\label{fig:susy_excl}
\end{figure}

As an illustrative example, we choose the case of a slepton decaying 
into a $b$ and a light quark. 
The low energy bounds only imply 
\[
\lambda'_{113} < 0.021 \; \frac{m_{\tilde f}}{100 \gev} ,
\]
will all other relevant $\lambda'_{ij3}$s allowed to be larger.
Thus we may, safely, choose $\lambda' = 0.02$ over the entire range 
of interest. In Fig.\ref{fig:susy_excl}, we present the 
discovery contours that are obtained by combining the results 
of our simulation (Fig.~\ref{fig:slep_b}$b$) with the branching 
fractions of Fig.~\ref{fig:br}$a$. 

The parameter space bounds as obtained from the other modes 
discussed in this paper are very similar. The case of the slepton
decaying into two non-$b$ quarks would, typically, lead to
constraints slightly weaker than those of Fig.\ref{fig:susy_excl}. 
This is particularly true for smaller values of $m_ {\tilde l}$. However,
for slepton masses close to the kinematic limit, $b$-tagging 
actually reduces the sensitivity.
Bounds derived from squark production, on the other hand, would 
depend crucially on the gluino mass. For small squark masses, 
decays into gluinos are unlikely and hence the bounds would 
be stronger than the ones described by Fig.\ref{fig:susy_excl}. 
For large squark masses, on the other hand, the gluino channel 
might open up thus reducing the \rpv\ branching fraction drastically 
(see Figs.\ref{fig:br}) and thereby decreasing the efficacy of the
search strategy discussed here.

Similar analyses obtain for leptoquarks and diquarks as well. In fact,
the case of a generic diquark is almost exactly the same as that for the 
slepton, but for the modification in the cross section on account of 
a possibly differing electric charge. A generic leptoquark, on the other 
hand, provides for an additional two-body decay channel namely 
into a quark and a neutrino. While such a mode precludes mass reconstruction,
it certainly can be used to improve the signal to noise ratio.

\section{Conclusions}
	\label{sec:concl}
To summarise, we discuss scalar particle pair production 
at a photon photon collider and their subsequent 2-body decays through 
$L$ violating interactions.
 
We find that the use 
of a few well-chosen kinematic cuts can eliminate the SM backgrounds to a 
considerable extent. Mass reconstruction is possible and is quite accurate 
almost over the entire kinematic range. For pair-production close to the 
threshold, the use of polarized electron and laser beams (used to obtain 
the high energy photons) can increase the production cross-sections manifold
without significantly altering the SM background. 

As the production cross-section is completely model-independent, 
the signal strength is a very good measure of the branching 
fractions into these $L$-violating decay modes. Consequently, bounds on the 
parameter space can be obtained, and we have done so for the case of
$R$-parity violating supersymmetric models. Although slepton production cross 
sections are the largest, the corresponding backgrounds are larger too. 
This renders the discovery reach for this channel to be somewhat
worse than that for up-type squarks decaying through a similar $LQD$ 
coupling. However, if the slepton were to decay into a $b$-quark, then 
$b$-jet identification could be used to increase the sensitivity 
significantly and make it competitive with the up-squark channel. 

It must be noted though that even if the $L$ violating
2-body decay modes be small, the daughter particles from the 
$L$-conserving channels will finally decay through a 
$L$-violating mode. Such a cascading process will leave its own 
tell tale signature. An analysis, albeit a more complicated one,
of the same would only serve to complement the present one. The 
discovery/exclusion plots presented here are thus only conservative 
ones and can be improved. 

Finally, a study of $B$ violating decays is but a straightforward 
extension of that presented in Section~\ref{sec:slepton} of this article.

\newcommand{\plb}[3]{{Phys. Lett.} {\bf B#1} (#3) #2}		       %
\newcommand{\prl}[3]{Phys. Rev. Lett. {\bf #1} (#3) #2}	       %
\newcommand{\rmp}[3]{Rev. Mod.	Phys. {\bf #1} (#3) #2}             %
\newcommand{\prep}[3]{Phys. Rep. {\bf #1} (#3) #2}		       %
\newcommand{\rpp}[3]{Rep. Prog. Phys. {\bf #1} (#3) #2}             %
\newcommand{\prd}[3]{{Phys. Rev.}{\bf D#1} (#3) #2}		       %
\newcommand{\np}[3]{Nucl. Phys. {\bf B#1} (#3) #2}		       %
\newcommand{\npbps}[3]{Nucl. Phys. B (Proc. Suppl.) 
           {\bf #1} (#3) #2}	                                       %
\newcommand{\sci}[3]{Science {\bf #1} (#3) #2}		       %
\newcommand{\zp}[3]{Z.~Phys. C{\bf#1} (#3) #2}		       %
\newcommand{\mpla}[3]{Mod. Phys. Lett. {\bf A#1} (#3) #2}             %
\newcommand{\astropp}[3]{Astropart. Phys. {\bf #1} (#3) #2}	       %
\newcommand{\ib}[3]{{\em ibid.\/} {\bf #1} (#3) #2}		       %
\newcommand{\nat}[3]{Nature (London) {\bf #1} (#3) #2}	       %
\newcommand{\nuovocim}[3]{Nuovo Cim. {\bf #1} (#3) #2}	       %
\newcommand{\yadfiz}[4]{Yad. Fiz. {\bf #1} (#3) #2 [English	       %
	transl.: Sov. J. Nucl.	Phys. {\bf #1} #3 (#4)]}	       %
\newcommand{\philt}[3]{Phil. Trans. Roy. Soc. London A {\bf #1} #2  
	(#3)}							       %
\newcommand{\hepph}[1]{(electronic archive:	hep--ph/#1)}	       %
\newcommand{\hepex}[1]{(electronic archive:	hep--ex/#1)}	       %
\newcommand{\astro}[1]{(electronic archive:	astro--ph/#1)}	       %

\end{document}